\documentclass[prl,twocolumn,showpacs]{revtex4}

\usepackage[dvips]{graphicx}
\usepackage{dcolumn}% Align table columns on decimal point
\usepackage{bm}% bold math

\begin{document}

\title{Deterministic uncertainty}

\author{Bartosz Telenczuk}
\email{mucha@poczta.fm}
\affiliation{Institute of Physics, Wroclaw University of Technology,Wybrzeze Wyspianskiego
27, 53-227 Wroclaw, Poland}

\author{Miroslaw Latka}
\email{mirek@rainbow.if.pwr.wroc.pl}
\homepage{http://www.if.pwr.wroc.pl/~mirek}
\affiliation{Institute of Physics, Wroclaw University of Technology,Wybrzeze Wyspianskiego
27, 53-227 Wroclaw, Poland}

\author{Bruce J. West}
\email{Bruce.J.West@us.army.mil}
\affiliation{Mathematics Division, Army Research Office, P.O. Box 12211, Research
Triangle, NC 27709-2211, USA}

\date{June 9, 2003}

\begin{abstract}
Equations of motion with delays naturally emerge in the analysis of
complex biological control systems which are organized around biochemically
mediated feedback interactions. We study the properties of a Mackey-Glass-type 
nonlinear map with delay -- the deterministic part of the stochastic
cerebral blood flow map (CBFM)  recently introduced to elucidate
the scaling properties of cerebral hemodynamics. We point out the
existence of \emph{deterministic} and \emph{nondeterministic subspaces}
in the reconstructed phase space of delay-difference equations and
discuss the problem of detection of nonlinearities in time series
generated by such dynamical systems.
\end{abstract}


\maketitle
Attractor reconstruction is certainly one of the more fundamental
methods of analysis  of nonlinear dynamical systems \cite{ott93}. Time series prediction,
nonlinear filtering, chaos control and targeting of trajectories all
exploit the properties of phase space \cite{kantz97}. However, the
number of variables accessible to measurement is frequently
limited. In many cases even the dimensionality of the system is not
known. The fact that phase space may be reconstructed using the time
evolution of just a \emph{single} dynamical variable in multidimensional
system paves the way for applications of nonlinear methods in natural
sciences.

As early as 1985 Babloyantz \emph{et al}. demonstrated that certain
nonlinear measures, first introduced in the context of chaotic dynamical
systems, changed during slow-wave sleep \cite{babloyantz85}. The
flurry of research work that followed this discovery focused on the
application of nonlinear dynamics in quantifying brain electrical
activity during different mental states, sleep stages, and under the
influence of the epileptic process (for a review see for example \cite{bwest95}).
Despite various technical difficulties, the number of applications
of nonlinear time series analysis has been growing steadily and now
includes prediction of epileptic seizures \cite{lehnertz98}, the
characterization of encephalopaties \cite{stam99}, monitoring of
anesthesia depth \cite{widman00}, characteristics of seizure activity
\cite{casdagli97}, fetal ECG extraction \cite{schreiber95}, and
control of human atrial fibrillation \cite{ditto00}. 

We have recently studied middle cerebral artery blood flow velocity
in humans using transcranial Doppler ultrasonography (TCD) \cite{latka03b,west99}.
We found that scaling properties of time series of the axial flow
velocity averaged over a cardiac beat interval may be characterized
by two exponents. The short-time scaling exponent determines the statistical
properties of fluctuations of blood flow velocity in short-time intervals
while the Hurst exponent describes the long-term fractal properties.
To elucidate the nature of two scaling regions characteristic of \emph{healthy}
individuals we introduced a stochastic cerebral blood flow map (CBFM).
The deterministic part of the CBFM is reminiscent of the Mackey-Glass
differential equation originally introduced to describe the production
of white blood cells \cite{mackey77,glass88} so that we refer to
it as the \emph{Mackey-Glass map (MGM):}

\begin{equation}
x_{n}=x_{n-1}-bx_{n-1}+a\frac{x_{n-\eta}}{1+x_{n-\eta}^{10}}.\label{MGMap}\end{equation}
Herein we  investigate the detailed properties of the above
difference equation from the viewpoint of detection of nonlinearities
and determinism. One invariably faces this detection problem  when dealing with
short, noisy physiological time series.

It is well established that the presence of delays even in such dynamically
simple systems as first-order differential equations may lead to intricate
and multidimensional evolution. Therefore, we begin our analysis with
the attractor reconstruction \cite{packard80,takens81,sauer91}.
 
\begin{figure}[p]
\includegraphics{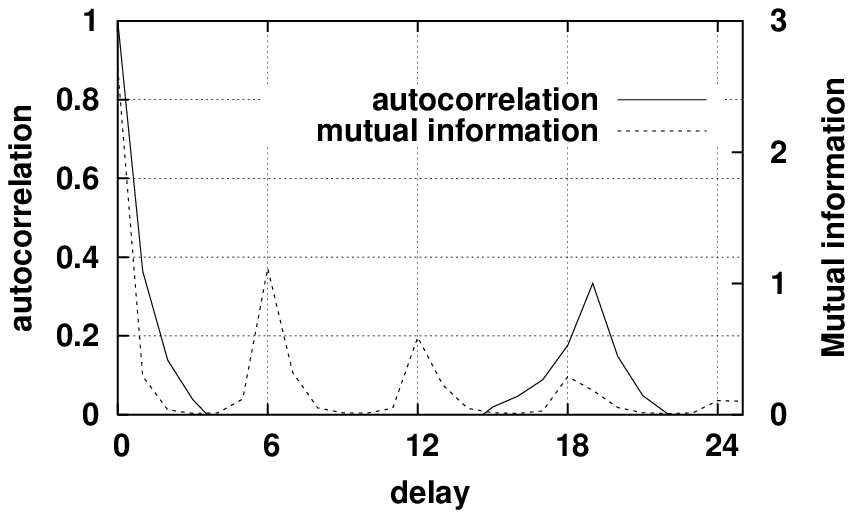}
\caption{Autocorrelation function (dashed line) and mutual information (solid
line) for the time series generated by the MGM (\ref{MGMap})
with the following parameters: $a=1.3$, $b=0.7$, $\eta=6$.\label{cap:AutocorrelationAndMutual}}
\end{figure}

 Let $\{ x_{i}\}_{i=1}^{N}$ be the time series generated by (\ref{MGMap}).
In the methods of delays the reconstructed trajectory is made up of
the following points:

\begin{equation}
\mathbf{X}_{n}=[x_{n-(m-1)J},x_{n-(m-2)J},...,x_{n-J,}x_{n}],\label{methodOfDelays}\end{equation}
where $J$ is the \emph{delay} or \emph{lag} and $m$
is the \emph{embedding dimension}. Thus, the reconstruction depends
upon two parameters which are not known a priori. Considerable
effort has been devoted to finding the best approach to selecting
judicious values of the lag $J$  see, for example, \cite{rosenstein93}  and references
therein. This is because for finite data sets, especially those
contaminated by noise, the choice of delay determines the outcome
of reconstruction. In Fig. \ref{cap:AutocorrelationAndMutual} we
compare the autocorrelation function and  mutual information of
the Mackey-Glass map time series ($a=1.3$, $b=0.7$, $\eta=6$). These
two methods are routinely used to estimate the proper reconstruction
lag. Typically $J$ is chosen as the delay for which the autocorrelation
function drops to a certain fraction of its initial value, e.g. $1/e$
\cite{albano88} or the first local minimum of the mutual information
\cite{fraser86}. Thus, we obtain the lag equal to 1 and 3, respectively.
It is apparent that the mutual information has distinct peaks at integer
multiples of the delay $\eta$ and therefore is particularly well
suited to detection of dynamical delays. The exponentially decreasing amplitude
of the consecutive maxima is a strong indication of sensitive dependence
on initial condition characteristic of chaotic dynamics so that we
proceed to calculations of the attractor dimension and the largest
Lyapunov exponent.

\begin{figure}
\includegraphics{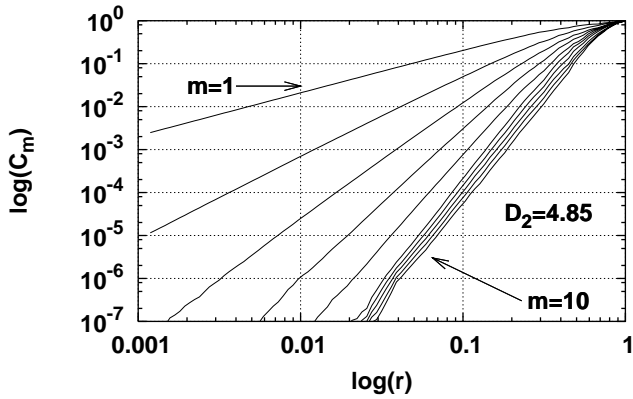}
\caption{\label{GP}Correlation integral $C_{m}$ for embedding spaces of
increasing dimension $m$ calculated for the MGM time series
used in Fig. \ref{cap:AutocorrelationAndMutual} \label{cap:Correlation-integral}.}
\end{figure}

Fig. \ref{cap:Correlation-integral} illustrates the process of calculation
of the correlation integral $C_{m}(r,J)$ for embedding spaces of
increasing dimension $m$ with $J=1$ \cite{grass83}. The slope of
the linear scaling regions in a double logarithmic plot converges
to the correlation dimension $D_{2}=4.85$. We obtain the same value
for $J=3.$ 

\begin{figure}
\includegraphics{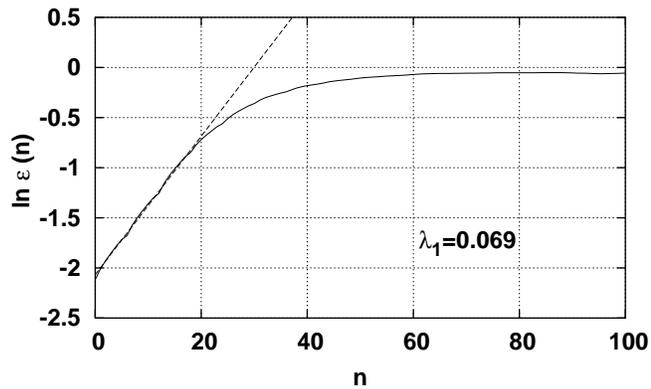}
\caption{Average divergence $\epsilon$ of initially close trajectories as
a function of number of iterations $n$ of the MGM ($a=1.3$,
$b=0.7$, $\eta=6$) is plotted in semilogarithmic scale. The slope
of the straight dashed line yields the value of the largest Lyapunov
exponent $\lambda_{1}.$ The reconstruction was done with $m=7$ and
$J=1.$\label{cap:Liapunov} }
\end{figure}

In Fig. \ref{cap:Liapunov} we graph the average divergence of initially
nearby trajectories as a function of the number of iterations of the
MGM \cite{rosenstein93}. In the semilogarithmic graph the slope
of the dashed straight line yields the value of the largest Lyapunov
exponent $\lambda_{1}=0.069$ which confirms that the underlying dynamics
is indeed chaotic. One can see in this plot that the average time
after which neighboring trajectories reach the boundaries of the available
phase space (as indicated by the constant value of their separation)
roughly coincides with the delay time interval where the maxima of
the mutual information are clearly pronounced \emph{cf}. Fig. \ref{cap:AutocorrelationAndMutual}.

Difference equations with delays such as the Mackey-Glass map are
seldom discussed in standard nonlinear dynamics textbooks. However,
let us revisit one of the three most fundamental nonlinear models
-- the Henon map:

\begin{eqnarray}
x_{n} & = & a-x_{n-1}^{2}+by_{n-1}\label{Henon}\\
y_{n} & = & x_{n-1}.\nonumber \end{eqnarray}
 From the second equation in (\ref{Henon}) we have $y_{n-1}=x_{n-2}$
and using this expression we arrive at:

\begin{equation}
x_{n}=a-x_{n-1}^{2}+bx_{n-2}.\label{delayedHenon}\end{equation}
 Hence, the two-dimensional Henon map is equivalent to the single difference
equation (\ref{delayedHenon}) with delay equal to 2. In the same
vein, we can rewrite (\ref{MGMap}) as a set of $\eta$ \emph{ordinary}
difference equations:

\begin{eqnarray}
x_{n}^{(1)} & = & (1-b)x_{n-1}^{(1)}+a\frac{x_{n-1}^{(\eta)}}{1+(x_{n-1}^{(\eta)})^{10}}\label{eq:set}\\
x_{n}^{(2)} & = & x_{n-1}^{(1)}\nonumber \\
x_{n}^{(3)} & = & x_{n-1}^{(2)}\nonumber \\
 & \vdots\nonumber \\
x_{n}^{(\eta)} & = & x_{n-1}^{(\eta-1)},\nonumber \end{eqnarray}
 which  may be interpreted as a perturbed \emph{periodic shift map:}

\begin{eqnarray}
x_{n}^{(1)} & = & x_{n-1}^{(\eta)}\nonumber \\
x_{n}^{(2)} & = & x_{n-1}^{(1)}\label{shiftMap}\\
 & \vdots\nonumber \\
x_{n}^{(\eta)} & = & x_{n-1}^{(\eta-1)}.\nonumber \end{eqnarray}
 This \emph{linear} map yields periodic orbits of length $\eta$ regardless
of the initial conditions. For brevity's sake, the evolution of a
difference equation with delay may be written with the help of the
shift operator $D_{f}$:

\begin{equation}
\mathbf{x}_{n+1}=D_{f}\mathbf{x}_{n}=[f(x_{n}^{(\eta)},x_{n}^{(1)}),x_{n}^{(1)},...,x_{n}^{(\eta-1)}]^{T}.\label{D}\end{equation}

\begin{figure}
\includegraphics{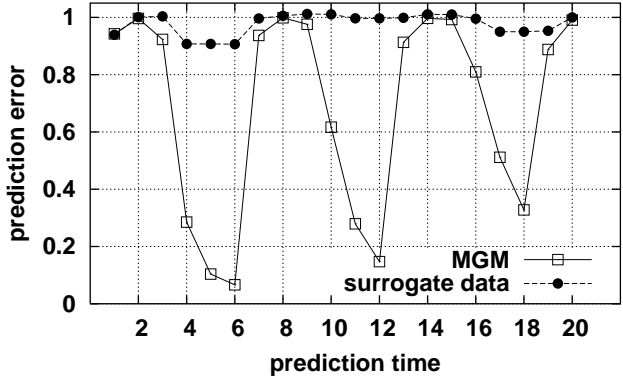}
\caption{The normalized forecasting error as a function of prediction time.
The simplest zeroth order phase-space algorithm was employed to predict
the values of the MGM time series and the corresponding surrogate
data. The embedding was done in three dimensions with the lag $J=1$.
The forecasting error was normalized by the standard deviation of the
time series.
\label{cap:Zeroth}}
\end{figure}

\begin{figure}
\includegraphics{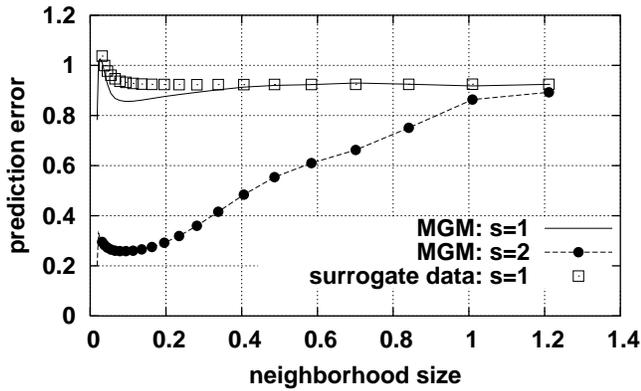}
\caption{The normalized prediction error as a function of the neighborhood
size used to construct a locally linear approximation of the MGM dynamics ($a=1.3$, $b=0.7$, $\eta=6$) or that of the surrogate
data. The embedding was done in three dimensions with the lag $J=2.$
The displayed curves correspond to the length $s$ of the prediction-time
interval\label{cap:Subspaces}}
\end{figure}

Now we turn the problem around and ask about the prospects of detecting
determinism in short, possibly corrupted with environmental noise,
time series generated by a difference equation with delay, such as
the MGM (\ref{MGMap}). Distinguishing chaos from environmental
noise has been a longstanding challenge to the life sciences where observed
time series usually have relatively few data points. Sugihara and May \cite{sugihara90}
pointed out that for chaotic dynamical systems the accuracy of the
nonlinear forecasting exponentially falls off with increasing prediction-time interval
at a rate related to the value of the Lyapunov exponent. On the other
hand, for uncorrelated noise the forecasting accuracy is roughly independent
of prediction interval. Their original idea of using short-term prediction
to detect determinism was later  extended \cite{kennel92,casdagli89,casdagli92,kantz97}.
In Fig. \ref{cap:Zeroth} we display the normalized prediction error
as a function of the prediction time. The simplest zeroth order phase-space
algorithm \cite{kantz97} was employed to forecast the values of the
MGM time series.  The forecasting accuracy not only initially
\emph{does improve} with the \emph{increasing} prediction-time interval
but also exhibits strong oscillations. For some relatively short forecasting
steps, the accuracy for the MGM time series is essentially
the same as that of the corresponding surrogate data (data which as
closely as possible share both the spectral properties and  distribution
with the original time series, but from which all nonlinear correlations
have been excised \cite{theiler92,schreiber00}). To shed some light
on this effect, let us assume that we forecast one step ahead ($s=1)$
and embed the MGM time series with $\eta=6$ in three dimensions ($m=3)$
with delay $J=2$ (for such delay we simply choose every other element
of the original time series to form three dimensional delay vectors
(\ref{methodOfDelays})). For clarity of presentation we disregard
the influence of the linear term in the shift operator $D_{f}$, \emph{i.e.}
$f(x_{n}^{(\eta)},x_{n}^{(1)})\approx g(x_{n}^{(\eta)})$. Using the
projection operator $P_{ijk}\mathbf{x}_{n}=[x_{n}^{(i)},x_{n}^{(j)},x_{n}^{(k)}]^{T}$,
we may describe the prediction in the reconstructed phase-space as:

\begin{eqnarray}
P_{135}\mathbf{x}_{n+1} & = & P_{135}(D_{g}\mathbf{x}_{n})\nonumber \\
 & = & P_{135}[g(x_{n}^{(6)}),x_{n}^{(1)},x_{n}^{(2)},...,x_{n}^{(5)}]^{T}\label{singleStep}\\
 & = & [g(x_{n}^{(6)}),x_{n}^{(2)},x_{n}^{(4)}]^{T}=\mathcal{F}_{1}(P_{624}\mathbf{x}_{n}).\nonumber \end{eqnarray}
 The propagator $\mathcal{F}_{1}$ depends on variables which have been \emph{discarded}
in the embedding process so that it should not come as a surprise
that accurate forcasting in this case  is  precluded since we selected
a \emph{nondeterministic subspace} with respect to the prediction-time
interval. However, if we choose to forecast two steps ahead ($s=2)$
then

\begin{eqnarray}
P_{135}\mathbf{x}_{n+2} & = & P_{135}D_{g}\mathbf{x}_{n+1}\nonumber \\
 & = & P_{135}[g(x_{n}^{(5)}),g(x_{n}^{(6)}),x_{n}^{(1)},...,x_{n}^{(4)}]^{T}\label{doubleStep}\\
 & = & [g(x_{n}^{(5)}),x_{n}^{(1)},x_{n}^{(3)}]^{T}=\mathcal{F}_{2}(P_{135}\mathbf{x}_{n})\nonumber \end{eqnarray}
 and the chosen subspace is closed with  respect to the propagator
$\mathcal{F}_{2}$. We refer to such subspace as \emph{deterministic.}
In Fig. \ref{cap:Subspaces} we verify our analysis by plotting the
normalized prediction error as a function of the neighborhood size
used to construct a locally linear approximation of the MGM
 dynamics \cite{casdagli92,hegger99}. It is apparent that for
$s=1$ the forecasting accuracy is hardly different from that of the
corresponding surrogate data. On the other hand, for $s=2$ the error
curve with a clearly pronounced minimum at small neighborhood sizes
is characteristic of nonlinear dynamical systems. The conclusions
we have drawn are valid also for  small to moderate linear coupling we so far
ignored by setting $f(x_{n}^{(\eta)},x_{n}^{(1)})\approx g(x_{n}^{(\eta)})$.
However, strong couplings may give rise to even more subtle effects
which we shall discuss elsewhere.

The forecasting error is one of many statistics applicable to testing
for nonlinearities in time series. For example, higher-order autocorrelations
such as

\begin{equation}
<(x_{n}-x_{n-d})^{3}>/<(x_{n}-x_{n-d})^{2}>\label{timeReversal}\end{equation}
may be used to measure time asymmetry, a strong signature of nonlinearity
\cite{kantz97}. Such autocorrelations are also helpful in pinpointing
another interesting property of delay-difference equations. Upon inspection
of (\ref{shiftMap}), one immediately realizes that the evolution
of the perodic shift map amounts merely to {}``shuffling'' of the
components of the state vector. However, \emph{random} shuffle of the
data is the crudest way of generation of surrogate data. If the dimensionality
of the system determined by the delay $\eta$ is sufficiently high,
it is not surprising that this type of time reversal test may not
detect nonlinearities. We generated the MGM time series ($a=1.3$,
$b=0.7$, $\eta=20$) along with 39 surrogate data sets and calculated
the corresponding value of (\ref{timeReversal}). For $11$ out of
$20$ delays $d$ taken from the interval $[1,20]$, we could not
reject the hypothesis that at the 95\% confidence level the MGM
time series results from a Gaussian linear stochastic process.

Delay-difference equations naturally emerge in the analysis of complex
biological control systems \cite{latka03b}. The unique properties
of such dynamical systems may lead to difficulties in recognizing
nonlinearities in their evolution especially when experimental time
series are short and and delays are long. To draw attention to this
issue we gave this paper a somewhat provocative title \emph{Deterministic
Uncertainty}.

\bibliographystyle{apsrev}
\bibliography{ghost}

\end{document}